\documentclass[]{article}

\usepackage[margin=0.75in]{geometry}
\usepackage[capposition=top]{floatrow}
\usepackage{xcolor}
\usepackage{amsmath}
\usepackage{amsfonts}
\usepackage{graphicx}
\usepackage{float}
\usepackage{hyperref}
\usepackage{setspace}
\usepackage{multicol,lipsum}
\usepackage{algorithm}
\usepackage{algpseudocode}

\title{Accounting for Missing Data in Public Health Research Using a Synthesis of Statistical and Mathematical Models}
\author{Paul N Zivich\textsuperscript{1}, 
	Bonnie E Shook-Sa\textsuperscript{2,3}, 
	Stephen R Cole\textsuperscript{1}, 
	Eric T Lofgren\textsuperscript{4}, 
	Jessie K Edwards\textsuperscript{1}}
\date{%
	\textsuperscript{1}Department of Epidemiology, Gillings School of Global Public Health, University of North Carolina at Chapel Hill, Chapel Hill, NC, USA\\%
	\textsuperscript{2}Department of Biostatistics, Gillings School of Global Public Health, University of North Carolina at Chapel Hill, Chapel Hill, NC, USA\\%
	\textsuperscript{3}Nuffield Department of Population Health, University of Oxford, Oxford, UK\\%
	\textsuperscript{4}Paul G. Allen School for Global Health, Washington State University, Pullman, WA, USA\\%
	~\\
	\today
}

\begin{document}
	
\maketitle

\begin{abstract}
	Introduction: Accounting for missing data by imputing or weighting conditional on covariates relies on the variable with missingness being observed at least some of the time for all unique covariate values. This requirement is referred to as positivity and positivity violations can result in bias. Here, we review a novel approach to addressing positivity violations in the context of systolic blood pressure. \\~\\
	Methods: To illustrate the proposed approach, we estimate the mean systolic blood pressure among children and adolescents aged 2-17 years old in the United States using data from 2017-2018 National Health and Nutrition Examination Survey (NHANES). As blood pressure was not measured for those aged 2-7, there exists a positivity violation by design. Using a recently proposed synthesis of statistical and mathematical models, we integrate external information with NHANES to address our motivating question. \\~\\
	Results: With the synthesis model, the estimated mean systolic blood pressure was 100.5 (95\% confidence interval: 99.9, 101.0), which is notably lower than either a complete-case analysis or extrapolation from a statistical model. The synthesis results were supported by a diagnostic comparing the performance of the mathematical model in the positive region. \\~\\
	Conclusion: Positivity violations pose a threat to quantitative medical research, and standard approaches to addressing nonpositivity rely on restrictive untestable assumptions. Using a synthesis model, like the one detailed here, offers a viable alternative.
\end{abstract}

\section*{Introduction}

Suppose we are interested in estimating the mean systolic blood pressure (SBP) among children and adolescents aged 2-17 in the United States (US) between 2017-2018. To estimate this mean, we use data from the National Health and Nutrition Examination Survey (NHANES), a nationally representative survey of children in the US \cite{noauthor_nhanes_nodate}.However, SBP measurements are partially missing. While we could restrict our data to those with measured SBP (i.e., conduct a complete-case analysis), this approach may be biased when there is a variable predictive of both SBP and missingness of SBP \cite{cole_missing_2023}. Instead, methods like weighting and imputation can be used to correct for missing outcome data \cite{vansteelandt_analysis_2010}. In absence of strong parametric modeling assumptions, these methods assume that SBP is measured for at least some people for each unique value of the measured variables. For example, if missingness depends on age (which is related to SBP) then these methods assume that SBP was a non-zero probability of being measured for each age. This assumption is referred to as positivity and can lead to bias when it is violated \cite{cole_missing_2023, westreich_invited_2010, petersen_diagnosing_2012, zivich_positivity_2022}. Existing approaches to address nonpositivity require modifying the scientific question or extrapolating from a statistical model. In recent work, Zivich and colleagues introduced a new approach that avoids the issues of these existing approaches \cite{zivich_transportability_2024, zivich_synthesis_2025}. Here, we review the problem posed by nonpositivity and illustrate this novel alternative with publicly available data and corresponding code in R and Python to replicate the analyses.

\section*{Methods}

Here, we consider the following variables collected by NHANES. The variable of interest, SBP (mm Hg), was measured up to three times. In this analysis, we took the mean of the available SBP measurements and SBP was set to missing if fewer than two measurements were available. Sampling weights were applied for inference to the US population. Additional covariates included age (years), height (centimeters), weight (kilograms), and gender (male, female). 

In the NHANES data, SBP was missing for 44\% of children. Unless SBP and the probability of missing SBP are independent, a complete-case analysis (i.e., restricting to only those with measured SBP) can be substantially biased \cite{cole_missing_2023}. To help make these ideas more precise, we introduce some notation. Let $X$ be age, $Y$ be SBP, and $R=1$ indicate that SBP was observed. Our parameter of interest, the mean SBP for children and adolescents, can be expressed as $\mu = E[Y]$, where $E[\cdot]$ is the expected value function. However, SBP is only observed forthose with $R=1$. For a complete-case analysis to be unbiased, SBP and missingness must be marginally independent or exchangeable \cite{cole_missing_2023}, which can be expressed as $E[Y] = E[Y \mid R=1]$ and indicates that those with a measured SBP are a random sample of all children and adolescents. This assumption would be valid if NHANES randomly determined which participants had their SBP measured. However, this assumption is not consistent with the NHANES design.

With missing data, a marginal exchangeability assumption is often unreasonable, as investigators rarely have such control over non-response. However, weaker assumptions regarding the missing data mechanism can be made. Rather than \textit{marginal} exchangeability, we can instead assume \textit{conditional} exchangeability. For didactic purposes, consider if missingness was independent of SBP conditional on age. The conditional exchangeability assumption can then be expressed as $E[Y \mid X=x] = E[Y \mid X=x, R=1]$ for all ages $x$ from 2-17. This assumption states that those with a measured SBP are a random sample \textit{withing each stratum of age} from 2 to 17, allowing those with a measured SBP to `stand-in' for those with an unmeasured SBP that are the same age. Implicitly, this exchangeability assumption comes with the positivity assumption that SBP has a non-zero probability of being measured for each unique age in our population \cite{zivich_positivity_2022, hernan_estimating_2006}. This assumption can be written as $\Pr(R=1 \mid X=x) > 0$ for all ages $x$ from 2-17. To illustrate why positivity is important, note that if everyone of a given age does not have a measured SBP, then there would be no one with a measured SBP to stand-in for them.

Given conditional exchangeability with positivity, we can express the parameter of interest, $\mu$, in terms of data we observed. Specifically, we can show that 
\begin{equation*}
	E[Y] = \sum_x E[Y \mid X=x, R=1] \Pr(X=x)
\end{equation*}
following from the law of total expectation over all ages $x$ from 2-17 and exchangeability with positivity. As the right-hand side is expressed in terms of the observed data, this result provides a recipe for estimating $\mu$. Specifically, it motivates a direct maximum likelihood estimator, also referred to as g-computation or imputation \cite{cole_missing_2023, vansteelandt_analysis_2010}. To apply this method,one first estimates a regression model for SBP given age among all observations with complete data and the NHANES sampling weights. Then using this estimated model, one predicts SBP given age for \textit{all} children in the data set. Effectively, this approach fills in the missing values for all observations. The mean SBP is then estimated by taking the sample-weighted average of the predicted SBP.

\subsection*{Nonpositivity}

Positivity is violated in our example as NHANES did not measure SBP for any participants $<8$ years old. To reiterate why nonpositivity is an issue, suppose we considered the following model for SBP given age
\begin{equation*}
	E[Y \mid X, R=1; \beta] = \beta_2 I(X=2) + \beta_2 I(X=3) + \cdots + \beta_{17} I(X=17) 
\end{equation*}
where $I(\cdot)$ is the indicator function. In this model, there is a unique coefficient for each age (i.e., the model is saturated), so no parametric modeling assumptions are imposed. Nonpositivity means that there is no one with SBP measured for ages $<8$, so the coefficients from $\beta_2$ through $\beta_7$ cannot be estimated with the NHANES data.

\subsection*{Existing Approaches}

To proceed when there is nonpositivity, we could modify the research question. Here, we could consider estimating the mean SBP in children aged 8-17 instead, represented mathematically as $E[Y \mid X \ge 8]$. While valid, this revision addresses a different research question and thus may limit the utility of this data analysis if interest is in the 2-17 age range. Instead, consider another approach to address the original research question. One could incorporate structural or parametric assumptions into a statistical model and use that model to extrapolate to ages with unmeasured SBP. For example, we might consider the following linear model
\begin{equation*}
	E[Y \mid X, R=1; \beta] = \beta_0 + \beta_1 X
\end{equation*}
which imposes a linear relationship between age and SBP. This model allows us to fill-in missing SBP values of those aged 2-7 using a model fit to those aged 8-17. Use of this regression model requires two important assumptions. First, the model assumes that there is a linear relationship between SBP and age among those 8-17. This assumption is possible to assess and relax with available data. Second, the use of this model assumes that the linear relationship extends to those aged 2-7. In other words, the model fit to those aged 8-17 is used to extrapolate to those aged 2-7. The assumption licensing this extrapolation cannot be assessed with the available NHANES data.

\subsection*{Proposed Approach}

To avoid the restrictive modeling assumptions of the extrapolation approach while still addressing the motivating question, we review a method based on a synthesis of statistical and mathematical models \cite{zivich_transportability_2024, zivich_synthesis_2025}. To motivate this approach, first note that we can divide the data into two parts: one where positivity is met and one where positivity is not met. As shorthand, let $X^* = 1$ if a NHANES participant's age was between 8-17 and $X^*=0$ otherwise. Using this additional notation and the idea of dividing the data regarding whether positivity is met, we can rewrite $\mu$ as 
\begin{equation*}
	E[Y] = E[Y \mid X^* = 1] \Pr(X^* = 1) + E[Y \mid X^* = 0] \Pr(X^* = 0)
\end{equation*}
which is simply a weighted average of the means in the positive and nonpositive regions. Since $X$ is observed for all participants, $\Pr(X^* = 1)$ and $\Pr(X^* = 0)$ can be directly estimated from the data. Therefore, how $E[Y \mid X^* = 1]$ and $E[Y \mid X^* = 1]$ can be learned in the remaining task. The core idea of the synthesis model is to rely on a statistical model to address missingness in the region with positivity (i.e., $E[Y \mid X^* = 1]$) and use a mathematical model to fill-in missing values using external information in the nonpositivity region (i.e., $E[Y \mid X^* = 0]$). 

Starting with $E[Y \mid X^* = 1]$, recall that this is the region with positivity which means standard statistical methods for missing data can be applied. Specifically, we can rely on those previous conditional exchangeability with positivity assumptions limited to those aged 8-17. These assumptions allow us to use the previously described g-computation procedure limited to those aged 8-17. Here, we use the following saturated model
\begin{equation*}
	E[Y \mid X, R=1, X^*=1; \beta] = \beta_8 I(X=8) + \beta_9 I(X=9) + \cdots + \beta_{17} I(X=17) 
\end{equation*}
which places no parametric constraints on the relationship between age and SBP for those aged 8-17. Unlike the extrapolation approach for dealing with nonpositivity, this statistical model does not require there to be a linear relationship between age and SBP for those ages 8-17.

Now we turn attention to $E[Y \mid X^* = 0]$. To reiterate, SBP was never measured for $X^* = 0$ in the data. Therefore, we need to look outside the available NHANES data to make progress. To start, one might consider plugging in extreme but still plausible values for the mean among 2-7 year olds (e.g., 70 to 120 mm Hg) \cite{zivich_synthesis_2025}. Plugging in these values provides a range of plausible values for $\mu$, often referred to as bounds \cite{manskiPartialIdentificationMissing2005}. These bounds provide a simple sensitivity analysis that only relies on the assumption that the true $E[Y \mid X^* = 0]$ lies somewhere within the chosen range. How $\mu$ changes across the range of plausible values can also be plotted to further explore the bounds.

When additional external information is available, we can move beyond a sensitivity analysis and instead build a mathematical model to fill-in the missing values. Here, values for those aged 2-7 are imputed based on published SBP distributions for US children and adolescents  \cite{flynn_clinical_2017}. The imputed values from the mathematical model come from random draws of age, gender, and height percentile specific SBP distributions. Here, we assume normal distributions under the assumption that the mean is equal to the median and the standard deviation is approximated from the 90\textsuperscript{th} percentile. Use of this mathematical model to fill-in missing values of SBP is justified by the assumption that the given external information is accurate for our context. Here, accurate external information for our context means that the source population for the published SBP descriptive statistics and NHANES population are similar for all other predictors of SBP besides age, gender, and height. This assumption would be violated, for example, if weight were distributed differently across data sources. As NHANES and the published SBP distributions are both intended to be representative of the US population, we expect any differences to be small.

To apply the synthesis model while also accounting for variability in both the statistical and mathematical models, we use a resampling procedure that incorporates the NHANES sample weights. The algorithm for this procedure is described in Appendix 1. Briefly, the NHANES data is resampled with replacement. The parameters of the statistical model are re-estimated and used to estimate $E[Y \mid X^* = 1]$. Similarly, new imputed values are drawn from the mathematical model in order to estimate $E[Y \mid X^* = 0]$. These values are then used to estimate $\mu$. This process is repeated a large number of times (i.e., 10,000), and the different estimates of $\mu$ are summarized by the median as the point estimate, and 2.5\textsuperscript{th} and 97.5\textsuperscript{th} percentiles as the 95\% confidence interval (CI). This procedure accounts for the sampling uncertainty in NHANES as well as the uncertainty in the mathematical model parameters. However, this procedure does not account for clustering in the NHANES design, so this procedure likely is an underestimate of the uncertainty.

As mentioned, the validity of this approach is premised on the mathematical model being built from accurate external information. While there are substantive reasons to believe this assumption to be reasonable, in this example the external information allows us to indirectly check the validity of this assumption. Here, the external information also included SBP distributions for those aged 8-17. Using this additional information, we compare the age-specific SBP values for those aged 8-17 between the statistical and mathematical models. To incorporate uncertainty, the resampling procedure is used to compare the difference between the age-specific SBP averages by model. While we believe such a check to have utility, it should be recognized that this diagnostic cannot assess the applicability of the mathematical model for the nonpositive region.

In practice, researchers may wish to incorporate more than one variables in their imputation model. For example, one could instead assume exchangeability conditional on age, gender, height, and weight within the positive region. In Appendix 2, we show how the synthesis approach can be applied with a parametric statistical model to account for these variables in the positive region. As alluded to earlier, use of parametric models relies on the assumption that the model is correctly specified. When this assumption is not met, estimates of $\mu$ may be biased. In Appendix 3, we describe an augmented inverse probability weighting estimator for synthesis models that relies on less restrictive parametric modeling assumptions. 

\subsection*{Research Ethics}

The design of NHANES was originally approved by the National Center for Health Statistics Ethics Review Board. The publicly available version of NHANES is de-identified and thus secondary analyses are not considered to be human subject research under United States federal regulation 45 CFR 46.102(e)(1).

\section*{Results}

\begin{figure}
	\centering
	\caption{Systolic blood pressure by age from observations in the 2017-2018 National Health and Nutrition Examination Survey and projections from the mathematical model}
	\includegraphics[width=0.6\linewidth]{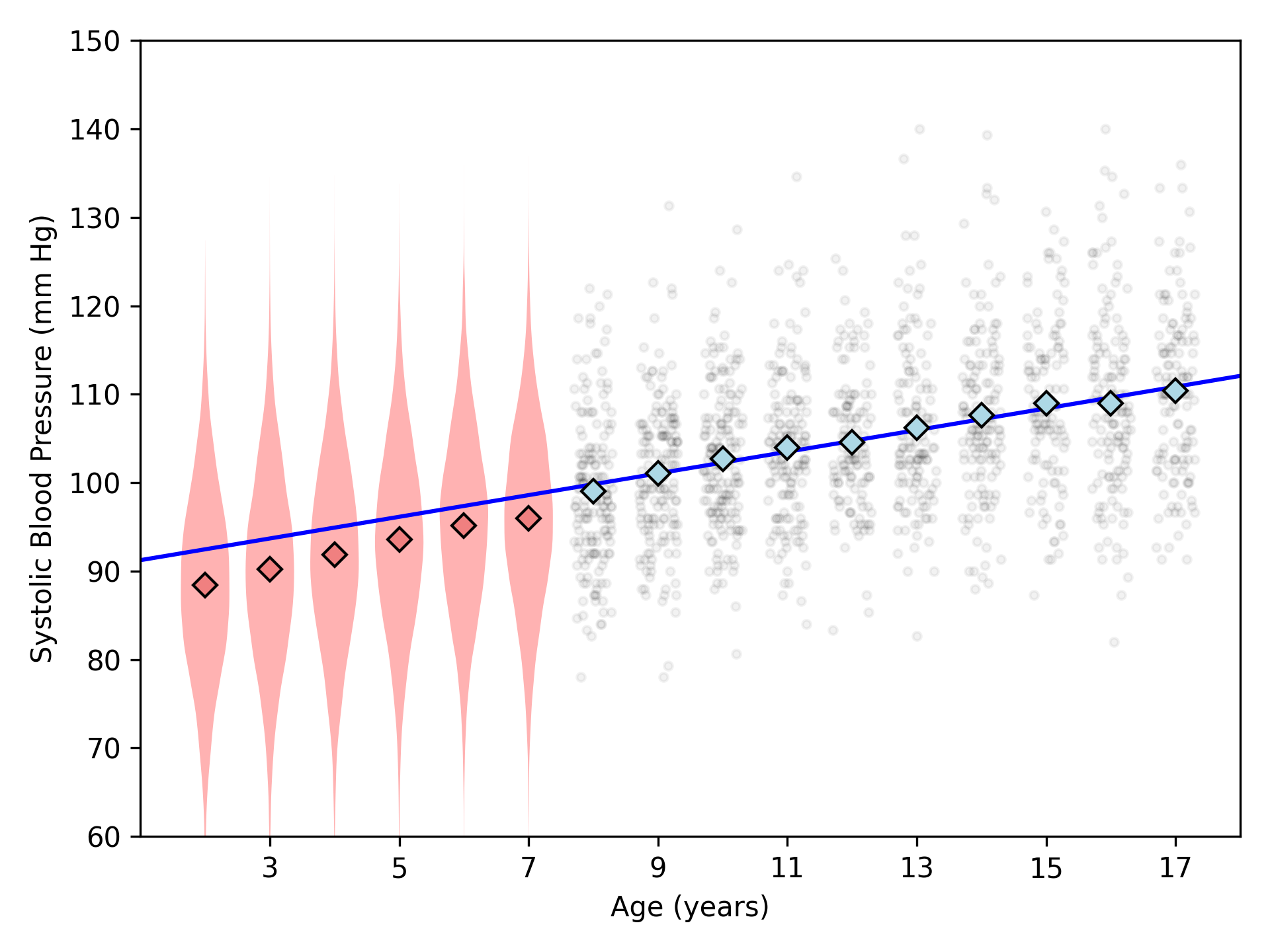}
	\floatfoot{Gray dots indicate the observed systolic blood pressure values by age (uniformly jittered for visualization) from NHANES. Systolic blood pressure was not measured for those younger than 8 years old. The blue diamonds indicate the age-specific means. The blue line indicates a linear model fit to the observed data and projected for those under 8. The red shaded region indicates the distributions of 20,000 simulated observations from the mathematical model, with the red diamonds indicating the corresponding mean. }
	\label{Figure1}
\end{figure}

\begin{figure}
	\centering
	\caption{Sensitivity analysis for non-positivity on the mean systolic blood pressure in the 2017-2018 National Health and Nutrition Examination Survey}
	\includegraphics[width=0.6\linewidth]{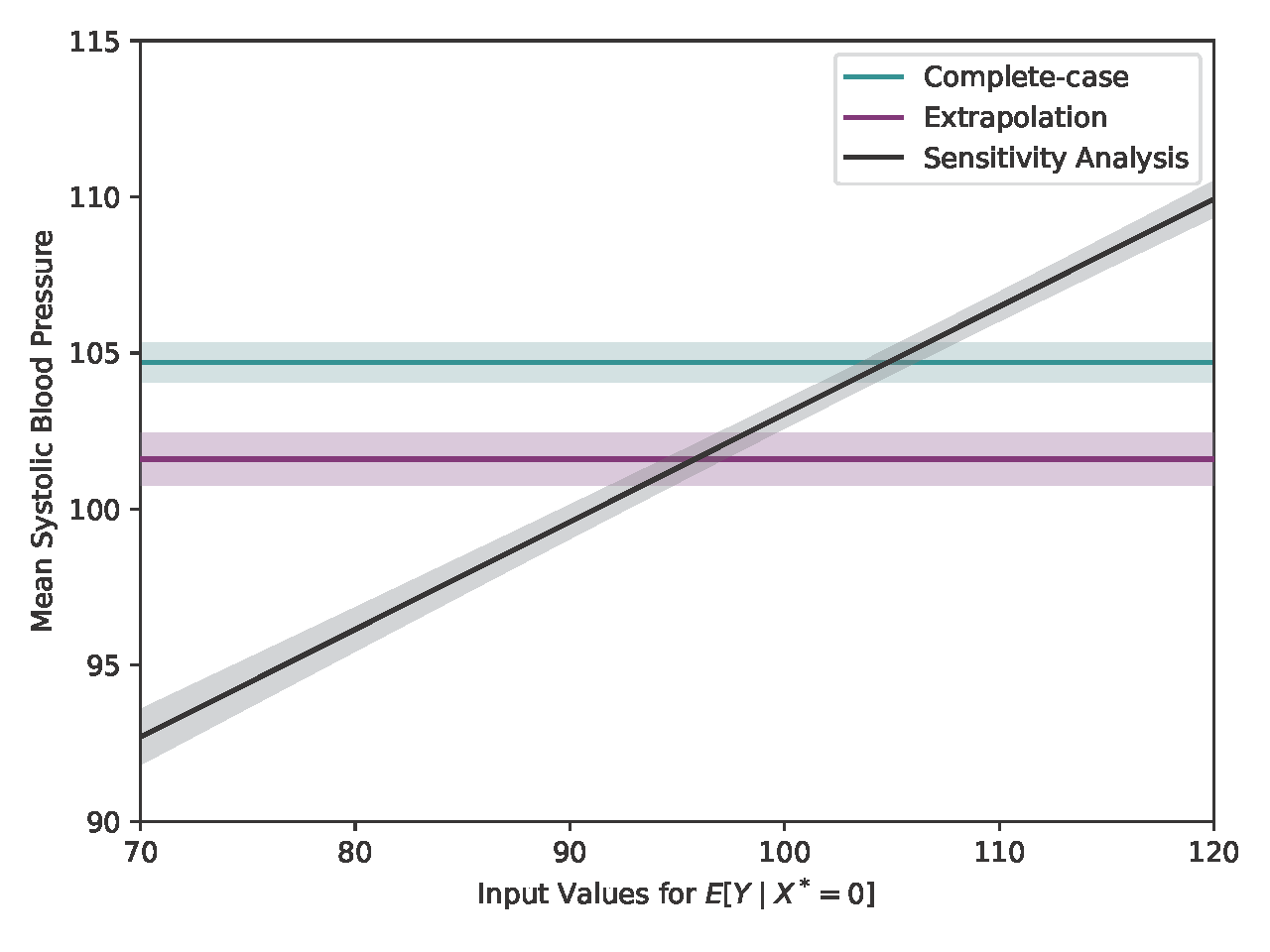}
	\floatfoot{The black line indicates how the mean systolic blood pressure for children aged 2-17 (y-axis) varies under differing input values for the mean in the non-positive region (i.e., children aged 2-8) across the range 70 to 120 mm Hg (x-axis). The colored horizontal lines represent the results from the complete-case and extrapolation analyses and show how they relate to the sensitivity analysis. Shaded regions denote the 95\% confidence intervals.}
	\label{FigureBounds}
\end{figure}

Those aged 2-7 made up 34\% of the NHANES weighted sample (Appendix Table \ref{Table1}). Among those aged 8-17, SBP was missing for 8\%. In a complete-case analysis, the estimated mean SBP was 104.7 (95\% CI: 104.1, 105.3). These results are questionable given what is known about the missing data and the relation between age and SBP. This estimate is more reflective of the mean SBP among those aged 8-17. With a linear extrapolation from the statistical model (Figure 1), the estimated mean SBP among children and adolescents was noticeably lower (mean: 101.6, 95\% CI: 100.8, 102.4). The difference between the complete-case analysis and extrapolation aligns with expectations, as those under 8 year old are expected to have lower SBP. If the mean SBP for those aged 2-7 was bound between 70-120, the bounds for the overall mean SBP would have ranged from 92.7 to 109.9 (95\% CI: 91.9, 110.5). Figure \ref{FigureBounds} shows how $\mu$ varies across this range. With a mathematical model, the estimated mean SBP was even lower than the extrapolation approach, at 100.5 (95\% CI: 99.9, 101.0). This difference between approaches arises from the linear model imputing higher mean SBP for lower ages relative to the mathematical model (Figure \ref{Figure1}). While the extrapolation and synthesis model results may not appear substantially difference, the observed 1.1 mm Hg difference is 2.9 times the estimated standard error of the extrapolation approach. When incorporating gender, height, and weight into the outcome model, results for the synthesis model were unchanged (Appendix 2). However, the extrapolation approach results moved closer to the synthesis approach. Results did not substantially change further when using an augmented inverse probability weighting estimator (Appendix 3).

\begin{figure}[H]
	\centering
	\caption{Comparison of statistical and mathematical model values for systolic blood pressure by age in the positive region.}
	\includegraphics[width=0.9\linewidth]{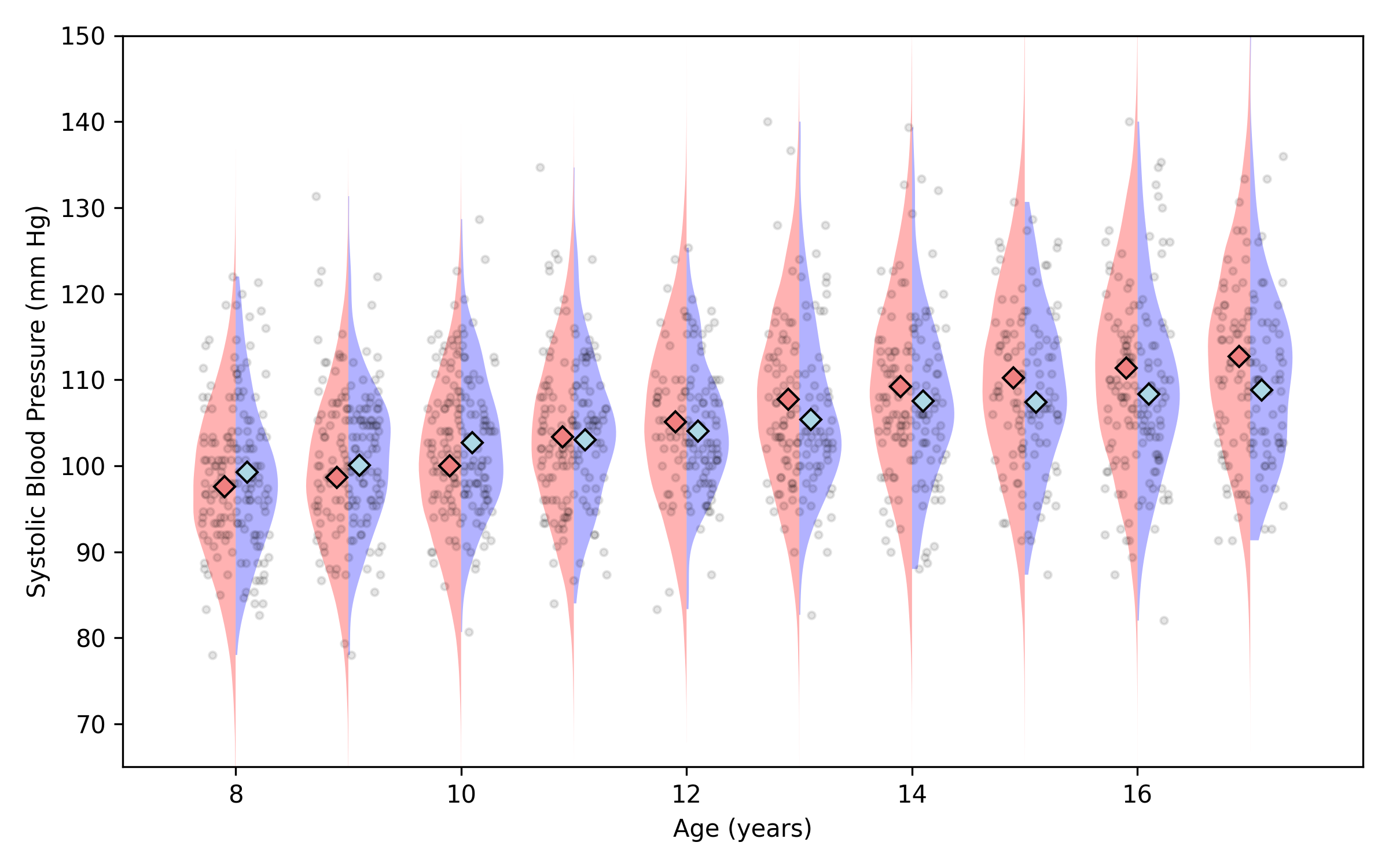}
	\floatfoot{Gray dots indicate the observed systolic blood pressure values by age (uniformly jittered for visualization) from NHANES. Blue diamonds indicate age-specific means from the statistical model. The red shaded region indicates the distributions of 20,000 simulated observations at each age from the mathematical model, with the red diamonds indicating the corresponding mean. }
	\label{Figure2}
\end{figure}

\begin{figure}[H]
	\centering
	\caption{Differences between mean systolic blood pressure comparing statistical versus mathematical models and corresponding 95\% confidence intervals.}
	\includegraphics[width=0.9\linewidth]{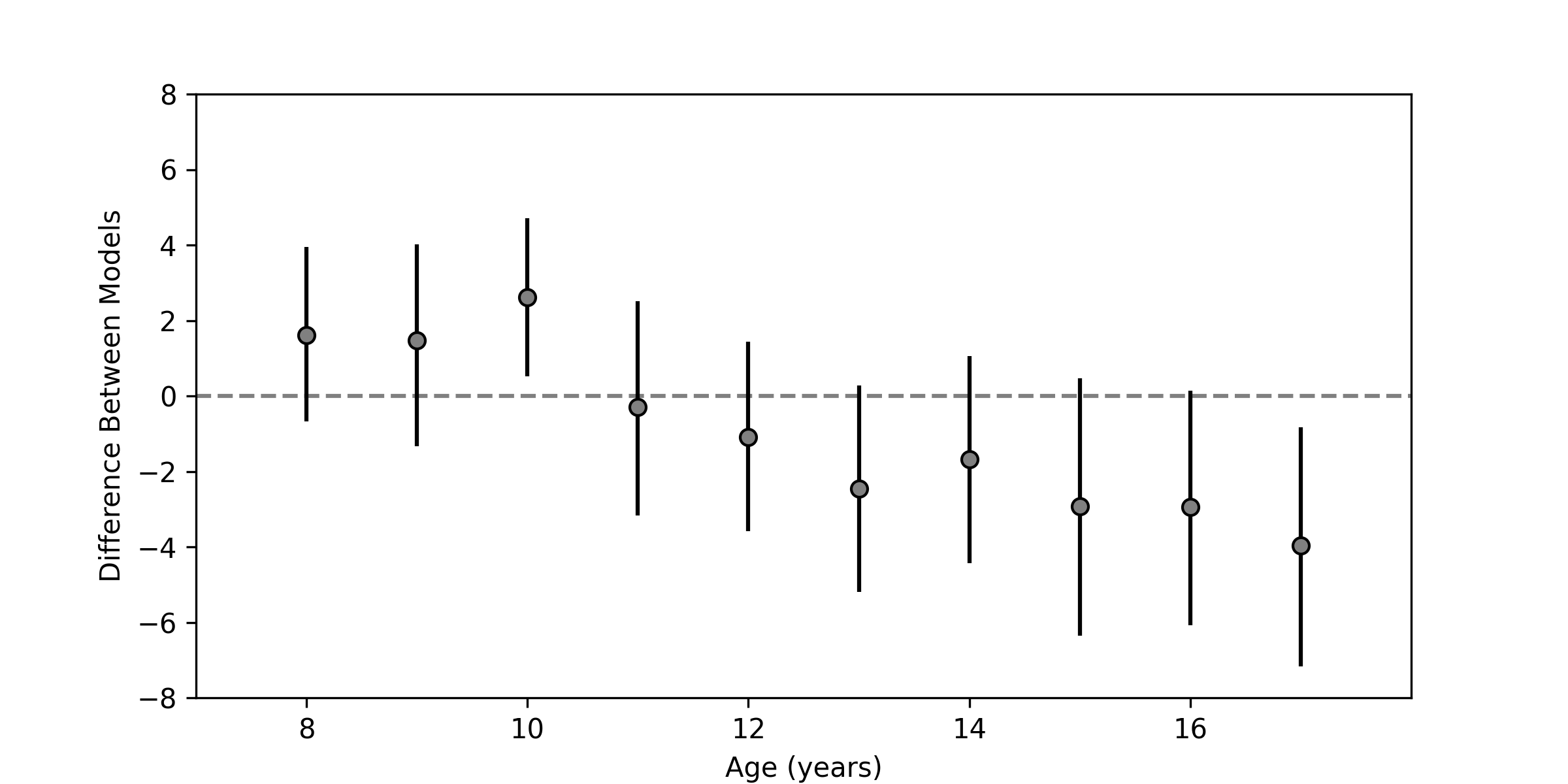}
	\floatfoot{Circles indicate the difference between the statistical model and mathematical model for each of the means among those in the positive region. Vertical lines indicate the corresponding 95\% confidence intervals.}
	\label{Figure3}
\end{figure}

When examining the validity of the mathematical model using the aforementioned diagnostic, the distribution of predicted SBP from the mathematical model overlaps with the observed SBP values for each age (Figure \ref{Figure2}). When comparing the differences between means and incorporating uncertainty, the estimated means from either approach were reasonably close to zero (Figure \ref{Figure3}). However, the statistical model results seemed to be consistently smaller than the mathematical model for those aged 15-17. Altogether we conclude that these results provide some support for the validity of the mathematical model in the positive region, which helps to support the belief that the mathematical model is also reasonable for the uncheckable nonpositive region.

\section*{Conclusions}

Nonpositivity threatens our ability to address important public health and clinical questions \cite{westreich_invited_2010}. While described here in the context of missing data, the positivity assumption also appears when addressing other biases, like confounding and selection bias \cite{zivich_positivity_2022}. Here, we illustrated an alternative in the context of missing data on SBP using publicly available data from NHANES. The proposed approach is based on integrating the available data with external information through a synthesis of statistical and mathematical models. Importantly, this approach does not require modifying the motivating question or rely on restrictive modeling assumptions, unlike competing methods.

There are a number of limitations to our analysis and the NHANES example provided here should be viewed only as illustrative. Regarding variance estimation, the approaches used here did not account for the use of cluster sampling in NHANES, so the variance may be underestimated. Extending the resampling algorithm to include additional complexities, like clustering, is an area for future research. Regarding the mathematical model, a fairly simple model was used here to directly impute SBP values. Simple models like this may not always be feasible. Instead, intermediate processes may need to be modeled. For example, imputing missing SBP data for adults on hypertensive medication with a mathematical model may instead involve modeling active drug concentrations and subsequent physiological responses \cite{heoQuantitativeModelBlood2016}. Designing and building complex mathematical models should follow best practices \cite{roberts_conceptualizing_2012, krijkamp_microsimulation_2018, slayton_modeling_2020} and evaluate diagnostic procedures whenever possible. Future work should illustrate use of these more advanced models. Finally, non-positivity may occur for more than a single variable. While the essential concepts of synthesis modeling still apply, illustrating their use in these contexts remains needed.

\section*{Acknowledgments}

Special thanks to Dr. Justin Lessler for discussion and suggestions on earlier versions of this work.
~\\~\\
Conflicts of Interest: None to declare. 
~\\~\\
Financial Support: This work was supported in part by the National Institutes of Health through K01AI177102 (PNZ), R01AI157758 (PNZ, BES, SRC, JKE), R01GM140564 (PNZ, JKE), K01AI182506 (BES), and R35GM147013 (ETL); and by Cancer Research UK through grant PRCRPG-Nov21/100001 (BES). The content is solely the responsibility of the authors and does not necessarily represent the official views of the National Institutes of Health or Cancer Research UK. The study sponsors had no role in the design, analysis, or decision to submit for publication.
~\\~\\
Data and Code: Data and code to replicate the example are available at \\ \url{https://github.com/pzivich/publications-code}.

\small
\bibliography{references}{}
\bibliographystyle{ieeetr}

\newpage
\normalsize

\section*{Appendix}

\setcounter{figure}{0}
\renewcommand{\thefigure}{A\arabic{figure}}
\setcounter{table}{0}
\renewcommand{\thetable}{A\arabic{table}}

\subsection*{Appendix 1: Resampling Algorithm for Uncertainty}

To incorporate the uncertainty in both estimation of the statistical model parameters and the chosen distributions for the mathematical model, the following resampling algorithm is used.

\noindent
\begin{enumerate}
	\item For $k$ from $1$ to $K$:
	\begin{itemize}
		\item[a.] Resample with replacement $n$ observations from the full NHANES data set.
		\item[b.] Divide the resampling data into the positive and non-positive regions.
		\item[c.] With the resampled NHANES data for the positive region, fit the specified statistical model using observations without missing data on the outcome. From the fitted model, predict the outcome for all observations in the positive region, regardless of whether observations had missing outcomes.
		\item[d.] With the resampling data for the non-positive region, randomly draw values from the imputed outcomes from the specified mathematical model. 
		\item[e.] Pool data from the positive and non-positive regions and take the mean of the imputed outcomes. If the data has sample weights, like NHANES, take the sample-weighted mean. Save this mean value as $\hat{\mu}_k$
	\end{itemize}
	\item Summarize the collection of $(\hat{\mu}_1, ..., \hat{\mu}_K)$. The point estimate can be obtained by taking the median. Two-sided 95\% confidence intervals can be obtained by taking the 2.5\textsuperscript{th} and 97.5\textsuperscript{th} percentiles. The values can also be plotted using a histogram to visualize the distribution.
\end{enumerate}

For intuition behind this procedure, note that this procedure involves re-estimating the statistical model parameters using data resampling with replacement, like the nonparametric bootstrap. Similarly, the mathematical model is used to draw values from the chosen distribution repeatedly. This provides a collection of estimates which are then jointly summarized.

When computing the confidence intervals for the bounds or sensitivity analysis, the mathematical model no longer contributes to the uncertainty since it is set to a fixed value. A resampling procedure (i.e., bootstrap) can be used for inference in these settings, or oe can use other appropriate variance estimation procedures (e.g., the empirical sandwich variance estimator).

\subsection*{Appendix 2: Extended Conditional Exchangeability}

Here, we consider the case where the conditional exchangeability assumption for missing data depends on more variables than age. Let $W$ denote the vector of the following variables: gender, weight, and height. Here, we make the assumption that missing data on SBP is non-informative given age, gender, weight, and height. The revised conditional exchangeability assumption is $E[Y \mid W=w. X=x] = E[Y \mid W=w, X=x, R=1]$ for all $w$ in the support of $W$ (i.e., all unique gender-weight-height combinations present in the population) and all ages $x$ from 2-17. Similar to before, this comes along with the following positivity assumption $\Pr(R=1 \mid W=w, X=x) > 0$ for all $w$ in the support of $S$ and all ages $x$ from 2-17. For the NHANES example, we assume that the positivity assumption holds except for children aged 2-7. This assumption matches the design of NHANES as the variables other than age did not preclude measurement of SBP. Given this exchangeability and positivity assumption, it follows that
\begin{equation*}
	E[Y] = E\left[ E(Y \mid W, X) \right] = E\left[ E(Y \mid W, X, R=1) \right]
\end{equation*}
which is a generalization of the identification result reported in the main paper using the law of iterated expectations and exchangeability with positivity by $W,X$. As before, positivity presents a challenge to identification and estimation.

As in the main manuscript, $E[Y]$ can be factored into the positive and non-positive regions since $E[Y] = E[Y \mid X^* = 1] \Pr(X^* = 1) + E[Y \mid X^* = 0] \Pr(X^* = 0)$. Again, a statistical and mathematical model are considered for $E[Y \mid X^* = 1]$ and $E[Y \mid X^* = 0]$, respectively. For $E[Y \mid X^* = 1]$, we can again rely on a modified version of the prior identification result to obtain
\begin{equation*}
	E[Y \mid X^* = 1] = E\left[ E(Y \mid W, X, X^* = 1, R=1) \mid X^* = 1 \right]
\end{equation*}
following conditional exchangeability with positivity by $W,X$ among those 8-17 years old. This result suggests a g-computation estimator where a model for $E(Y \mid W, X, X^* = 1, R=1)$ is fit and then used to fill-in the observations in the positive region. Unlike in the main manuscript, a saturated model is no longer as feasible to fit. Instead, a parametric model is used. Here, age is modeled using a restricted quadratic spline with knots at 10, 13, 15. Similarly, both weight and height were modeled using restricted quadratic splines with 3 knots each at 25, 35, 85 and 90, 140, 160 (inches); respectively. Models were separately estimated by gender.

For the mathematical model, no change is required in this extended version of the exchangeability assumption. By dividing the parameter into two pieces, the mathematical model solely depends on the external information available, which has not changed with the extension of exchangeability to include $W$ for the positive region. Therefore, the same mathematical model can be used as in the main paper, which already incorporates age, gender, and height.

For estimation with the synthesis model, the same resampling procedure described in Appendix 1 is used. For comparison with the synthesis model, we also estimated the mean SBP in children aged 8-17 accounting for informative missing by age, gender, weight, and height and extrapolating SBP for those aged 2-17. For both analyses, we drop the single additional observation with a missing value for weight.

When applying the extrapolation approach, the estimated mean SBP was 100.8 (95\% CI: 97.7, 103.8). This estimate is slightly lower than the extrapolation result reported in the main paper with noticeably wider confidence intervals. The synthesis model instead estimated a mean SBP of 100.5 (95\% CI: 99.9, 101.0), which matched the synthesis model results of the main paper up to the first decimal place. The synthesis model remaining similar with a change in the extrapolation approach indicates that the other variables (height, weight, gender) were not strongly related to missingness within the positive region but may have improved the extrapolations across the nonpositive region for the statistical model. Further, age was modeled more flexibility, which may have led to better agreement between models. This analysis further highlights the advantage of separating the parameter into positive and non-positive regions within the synthesis model.

\subsection*{Appendix 3: Alternative Synthesis Estimators for Missing Data}

In the section, we consider alternative statistical models for estimation of $\mu$. Continuing with the use of parametric models from the prior section, their use adds an additional assumption regarding correct statistical model specification. G-computation relies on modeling the outcome process, but inverse probability weighting (IPW) estimators rely on a statistical model for the missingness process \cite{cole_missing_2023}. Finally, augmented inverse probability weighting (AIPW) estimators use both statistical models and are doubly robust in that the estimator is consistent as long as one of the two statistical models is correctly specified \cite{vansteelandt_analysis_2010}. Here, we review the use of IPW and AIPW with synthesis estimators.

As in the main paper, the parameter of interest can be decomposed into
\begin{equation}
	E[Y] = E[X \mid X^* = 1] \Pr(X^* = 1) + E[X \mid X^* = 0] \Pr(X^* = 0)
	\label{EqA31}
\end{equation}
As before, a statistical model is used to estimate $E[X \mid X^* = 1]$ and a mathematical model is used to estimate $E[X \mid X^* = 0]$. In the following, we consider alternative estimators for $E[X \mid X^* = 1]$ but the mathematical model for $E[X \mid X^* = 0]$ remains the same since the external information is not modified.

\subsubsection*{Inverse Probability Weighting}

The IPW identification expression is
\begin{equation*}
	E[Y \mid X^* = 1] = E \left[ \frac{Y R}{\Pr(R=1 \mid X,W,X^*=1)} \mid X^* = 1 \right]
\end{equation*}
Notice that the IPW expression involves a model for the missingness process (denominator of the right-hand side) rather than the outcome process. Here, this probability can be estimated using logistic regression, or weighted logistic regression when there are sampling weights (as in the NHANES example). After estimating the probability of SBP being observed, the weighted mean of SBP among the complete cases can be computed. This estimate of $E[X \mid X^* = 1]$ can then be combined with the mathematical model following the expression in Equation \ref{EqA31}. For estimation of the uncertainty, one could then use the re-sampling procedure described in Appendix 1.

Here, the extrapolation approach is no longer straightforward to apply since the IPW estimator does not specify an outcome model from which to extrapolate. As described elsewhere \cite{zivich_transportability_2024, zivich_synthesis_2025}, using an IPW estimator to extrapolate requires the addition of an outcome model.

\subsubsection*{Augmented Inverse Probability Weighting}

The AIPW estimator combines g-computation and IPW estimators in such a way that if either the statistical model for the outcome process or the missingness process, but not necessarily both, is correctly specified then the AIPW estimator is consistent. Practically, this means that the AIPW estimator weakens the underlying parametric modeling assumptions relative to the g-computation and IPW estimators. There are several ways to implement AIPW \cite{bang_doubly_2005}, but here we describe a weighted-regression implementation due to its ease of implementation. To estimate $E[X \mid X^* = 1]$, first one estimate the probability of being observed, $\Pr(R=1 \mid X,W,X^*=1)$, using a (weighted) logistic regression model. These probabilities are then inverted, $R / \Pr(R=1 \mid X, W, X^* = 1)$, to construct inverse probability of missingness weights. Next, one fits a model for the outcome as done with g-computation. However, this model is not fit using weighted least squares regression with the inverse probability of missingness weights. If the study has sampling weights, this model is fit using the overall weight for an observation (computed as the product of the sampling weight and missingness weight). This fitted model can then be used following the g-computation procedure described in the main paper. This estimate of $E[X \mid X^* = 1]$ can again be combined with the mathematical model following the expression in Equation \ref{EqA31}. For estimation of the uncertainty, one can again use the re-sampling procedure described in Appendix 1.

Since an outcome model is specified, the extrapolation approach is straightforward to use with the AIPW estimator.

\subsubsection*{Results}

For the missingness model, a weighted logistic model was used including main effects and splines for the continuous variables (age, height, weight). Applying the AIPW estimator for the extrapolation approach gave similar results to the extrapolation g-computation (mean: 101.1; 95\% CI: 98.4, 103.8). Again, the synthesis results were unchanged for the reported number of decimals (mean: 100.5; 95\% CI: 99.9, 101.0).
~\\~\\~\\
\begin{table}[H]
	\caption{Age distribution and missing systolic blood pressure ($n=2572$)}
	\centering	
	\begin{tabular}{ccc}
		\hline
		Age & Number of Participants (\%)* & Number Missing SBP (\%) \\ \hline
		2   & 197 (5.3\%)                  & 197 (100\%)             \\
		3   & 157 (6.2\%)                  & 157 (100\%)             \\
		4   & 168 (6.1\%)                  & 168 (100\%)             \\
		5   & 166 (5.9\%)                  & 166 (100\%)             \\
		6   & 147 (5.7\%)                  & 147 (100\%)             \\
		7   & 153 (5.3\%)                  & 153 (100\%)             \\
		8   & 183 (6.6\%)                  & 15 (8.1\%)              \\
		9   & 190 (7.2\%)                  & 25 (13.2\%)             \\
		10  & 183 (6.2\%)                  & 16 (8.7\%)              \\
		11  & 168 (6.3\%)                  & 16 (9.5\%)              \\
		12  & 144 (6.1\%)                  & 15 (10.4\%)             \\
		13  & 143 (6.9\%)                  & 11 (7.7\%)              \\
		14  & 153 (6.9\%)                  & 11 (7.2\%)              \\
		15  & 127 (5.5\%)                  & 9 (7.1\%)               \\
		16  & 149 (7.2\%)                  & 6 (4.0\%)               \\
		17  & 144 (6.6\%)                  & 10 (6.9\%)              \\ \hline
	\end{tabular}
	\floatfoot{
		SBP: systolic blood pressure. \\
		* Sample-weighted percentage using the full sampling weights from the 2017-2018 National Health and Nutrition Examination Survey. }
	\label{Table1}
\end{table}

\end{document}